\documentclass[sigconf]{acmart}

\usepackage[ruled]{algorithm2e}
\usepackage{multirow}
\usepackage{threeparttable}
\usepackage{balance}

\copyrightyear{2024} 
\acmYear{2024} 
\setcopyright{acmlicensed}\acmConference[FPGA '24]{Proceedings of the 2024 ACM/SIGDA International Symposium on Field Programmable Gate Arrays}{March 3--5, 2024}{Monterey, CA, USA}
\acmBooktitle{Proceedings of the 2024 ACM/SIGDA International Symposium on Field Programmable Gate Arrays (FPGA '24), March 3--5, 2024, Monterey, CA, USA}
\acmDOI{10.1145/3626202.3637576}
\acmISBN{979-8-4007-0418-5/24/03}

\acmSubmissionID{\#359}

\begin{document}

\title{Table-Lookup MAC: Scalable Processing of Quantised Neural Networks in FPGA Soft Logic}

\author{Daniel Gerlinghoff}
\orcid{0000-0001-7332-1663}
\affiliation{%
  \institution{Institute of High Performance Computing, Agency for Science, Technology and Research}
  \streetaddress{1 Fusionopolis Way, \#16-16 Connexis}
  \country{Singapore}
  \postcode{138632}
}
\email{gerlinghoffd@ihpc.a-star.edu.sg}

\author{Benjamin Chen Ming Choong}
\orcid{0009-0007-9343-7517}
\affiliation{%
  \institution{Institute of High Performance Computing, Agency for Science, Technology and Research}
  \streetaddress{1 Fusionopolis Way, \#16-16 Connexis}
  \country{Singapore}
  \postcode{138632}
}
\email{choongbcm@ihpc.a-star.edu.sg}

\author{Rick Siow Mong Goh}
\orcid{0000-0001-9116-1595}
\affiliation{%
  \institution{Institute of High Performance Computing, Agency for Science, Technology and Research}
  \streetaddress{1 Fusionopolis Way, \#16-16 Connexis}
  \country{Singapore}
  \postcode{138632}
}
\email{gohsm@ihpc.a-star.edu.sg}

\author{Weng-Fai Wong}
\orcid{0000-0002-4281-2053}
\affiliation{%
  \institution{National University of Singapore}
  \streetaddress{Computing 1, 13 Computing Drive}
  \country{Singapore}
  \postcode{117417}
}
\email{wongwf@nus.edu.sg}

\author{Tao Luo}
\orcid{0000-0002-3415-3676}
\affiliation{%
  \institution{Institute of High Performance Computing, Agency for Science, Technology and Research}
  \streetaddress{1 Fusionopolis Way, \#16-16 Connexis}
  \country{Singapore}
  \postcode{138632}
}
\email{tluo001@e.ntu.edu.sg}
\authornote{Corresponding author}

\hyphenation{ImageNet Vivado LogicNets FPGA FPGAs}

\begin{abstract}
Recent advancements in neural network quantisation have yielded remarkable outcomes, with three-bit networks reaching state-of-the-art full-precision accuracy in complex tasks. These achievements present valuable opportunities for accelerating neural networks by computing in reduced precision. Implementing it on FPGAs can take advantage of bit-level reconfigurability, which is not available on conventional CPUs and GPUs. Simultaneously, the high data intensity of neural network processing has inspired computing-in-memory paradigms, including on FPGA platforms. By programming the effects of trained model weights as lookup operations in soft logic, the transfer of weight data from memory units can be avoided, alleviating the memory bottleneck. However, previous methods face poor scalability -- the high logic utilisation limiting them to small networks/sub-networks of binary models with low accuracy. In this paper, we introduce \textit{Table Lookup Multiply-Accumulate} (TLMAC) as a framework to compile and optimise quantised neural networks for scalable lookup-based processing. TLMAC clusters and maps unique groups of weights to lookup-based processing elements, enabling highly parallel computation while taking advantage of parameter redundancy. Further place and route algorithms are proposed to reduce LUT utilisation and routing congestion. We demonstrate that TLMAC significantly improves the scalability of previous related works. Our efficient logic mapping and high degree of reuse enables entire ImageNet-scale quantised models with full-precision accuracy to be implemented using lookup-based computing on one commercially available FPGA.
\end{abstract}



\maketitle

\section{Introduction}
\begin{figure}[t]
  \centering
  \includegraphics[width=\linewidth]{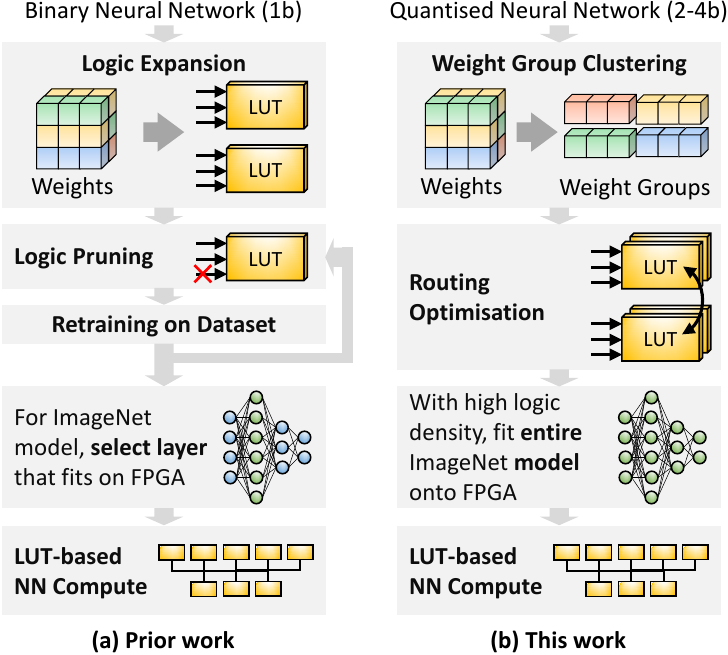}
  \caption{Comparison of steps involved in obtaining FPGA soft logic for neural networks. While prior research~\cite{wang2019lutnet, wang2022logic} customises and constrains the training process to achieve LUT compatibility, TLMAC derives optimised LUT initialisations from state-of-the-art quantised models directly.}
  \Description{Flow charts for the deployment of neural networks on soft logic using prior versus this method.}
  \label{fig: tlmac overview}
\end{figure}

FPGAs have emerged as an important platform for AI workloads, known for their ability to process many data streams concurrently at a high throughput and low latency. Furthermore, their malleability enables users to adapt and customise architectures to the ever-evolving landscape of algorithms. In view of their bit-level reconfigurability, FPGAs are especially suitable accelerators for \textit{quantised neural networks} (QNN), in which data precision is reduced to 2 to 7 bits. Recent quantisation methods have advanced to reach full-precision model accuracy with only 3-bit weights and activations~\cite{liu2022nonuniform}. Programmable logic allows FPGAs to implement low-bit arithmetic that is not available in CPU and GPU devices.

Numerous high-performance FPGA-based neural network accelerators have been proposed~\cite{venieris2018fpgaconvnet, liu2021toward, latotzke2022design}. A conventional design approach is to use BRAM blocks for storing weights and activation data, while DSP slices or logic fabric execute multiply-accumulate (MAC) operations which dominate neural network computation. However, as AI workloads are data-intensive, this separation between storage (BRAM) and processing units (DSPs) demands a high volume of data transfers between them~\cite{kabir2023making}. To address these inefficiencies, custom BRAM circuits for FPGAs supporting \textit{computing-in-memory} (CIM) were brought forward, enabling the BRAM to function as both a compute engine and traditional memory~\cite{arora2022comefa, chen2023bramac, wang2021compute}. However, these circuits were largely tested through simulations.

An alternative approach implements computing-in-memory functionality by leveraging the existing \textit{soft logic} resources (lookup tables, or LUTs)~\cite{umuroglu2020logicnets, wang2019lutnet}. Thereby, the activations are provided as LUT inputs, while the static model weights become part of the LUT initialisations. More specifically, the LUTs' logic functions yield the pre-calculated results of the MAC operations between a group of weights and all possible input combinations. This table lookup constitutes a memory access, while simultaneously performing computations. This avoids model weights from being transferred from BRAM or off-chip memory.

The process of converting neural networks to LUT-based computing is depicted in Figure~\ref{fig: tlmac overview}(a). To date, LUT-based computing works have targeted mainly \textit{binary neural networks} (BNN), in which both weights and activations are 1-bit values. Binary weights are first expanded into logic functions and pruned, followed by iterative retraining to restore accuracy loss.
However, current methods face two drawbacks that severely limit applicability:

\begin{enumerate}
    \item \emph{Significant logic resource requirements}. The neural networks obtained by these methods are fully unrolled, leading to high resource demands. As a consequence, only small BNNs are supported. Larger BNN models are needed for more complex tasks such as ImageNet classification. Unfortunately, only a single layer of these models could be deployed on an FPGA even after logic pruning~\cite{umuroglu2020logicnets}.
    
    \item \emph{Low prediction accuracy}. The constraint of using small BNNs and logic pruning to meet the resource budget causes much lower accuracy than the state-of-the-art. The accuracy gap remains large even with custom retraining.
\end{enumerate}

To address these challenges, we introduce \textit{table lookup multiply-accumulate} (TLMAC), a framework to compile quantised neural networks into table lookups in FPGAs. The TLMAC process is illustrated in Figure~\ref{fig: tlmac overview}(b). With low-bit quantisation, the number of unique weight values (and hence, weight patterns) is greatly reduced. Our framework encompasses weight group clustering which maps weights efficiently to lookup-based processing elements, achieving a high degree of weight reuse and parallelism. In addition, a routing optimisation algorithm is presented that arranges weight assignments between LUTs to reduce routing utilisation. As a result, TLMAC enables an entire QNN model for tasks as complex as ImageNet to be implemented in soft logic of an FPGA while maintaining original accuracy.

\section{Related Research}
\subsection{Computing-in-Memory on FPGAs}
In-memory computing has garnered attention for its ability to significantly reduce energy consumption of multiply-accumulate (MAC) operations, which are fundamental in neural networks. Frequently, CIM methods are developed at the circuit or even lower device level~\cite{wang2021memory, jung2022crossbar}. By requiring the fabrication of dedicated integrated circuits, however, those technologies are presently inaccessible for FPGA designers and machine learning engineers.

Some techniques take advantage of existing logic and routing infrastructure in FPGA devices and employ BRAM circuits that maintain the interface of conventional memory blocks. Compute-Capable Block RAMs~\cite{wang2021compute} adapts the BRAM to compute in-memory as proposed in~\cite{eckert2018neural}. To implement bit-serial operations close to the memory port, it had to add logic for serialisation and enhance memory cell access. CoMeFa~\cite{arora2022comefa} enhances both ports of dual-port BRAMs with write drivers and sense amplifiers. Circuits for processing elements are proposed that write input and read output data. Their bit-serial computation principle has also been applied to deep learning workloads~\cite{arora2023comefa}. BRAMAC~\cite{chen2023bramac} overcomes some of the previous implementation and performance restrictions by executing operations on separate dummy BRAMs. Two multiplications and one addition are achieved in a single memory operation, by using hybrid bit-serial/bit-parallel arithmetic. All these proposals require complex additional circuitry and are currently not commercially available.

\subsection{Table Lookup-based Computing}
Aside from block RAM, FPGAs contain a vast amount of lookup tables that can implement arbitrary Boolean functions on its binary inputs. Their smaller size and greater number make them a good alternative for BRAMs in CIM designs. Binary neural networks are particularly well suited as the neuron model uses \texttt{XNOR} operations that can directly be transferred onto LUTs after unrolling the layers~\cite{murovivc2019massively}. In LUTNet~\cite{wang2019lutnet}, pruning was applied and simple \texttt{XNOR} operations were transformed into Boolean functions with more inputs and higher complexity, significantly increasing the logic density. Logic Shrinkage~\cite{wang2022logic} took this further by eliminating LUTs of low importance. Despite these optimisations, however, scalability remains an issue and FPGA resources were only sufficient to accommodate a single AlexNet layer or ResNet block, respectively. Furthermore, the binarisation of weights and activations still results in a lower accuracy compared to few-bit neural networks.

LogicNets~\cite{umuroglu2020logicnets} developed a procedure for the training of quantised models whose parameters can be converted directly into a netlist. While they achieved a 15ns latency, only small models with around 1000 neurons were supported. PolyLUT~\cite{andronic2023polylut} generalises LogicNets by increasing the expressiveness of the model via more complex neuron functions. This complexity is absorbed into the LUT during inference time. Both methods are, however, only applicable to small problems due to LUT resource constraints.

Lastly, weightless neural networks incorporate lookup operations into the network architecture and training process, making them particularly suited for in-memory computing. The WiSARD network~\cite{aleksander2009brief}, for example, represents its neuron behaviour by means of a memory block. However, their excessive resource use obstructs deployment on commercial FPGA devices. In order to make the classification of MNIST images on a real-world FPGA feasible, Ferreira~et~al.~\cite{ferreira2019feasible} used hash tables which alleviated the exploding memory consumption. LogicWiSARD~\cite{miranda2022logicwisard} converts RAM contents into Boolean logic functions to reduce the use of arithmetic circuits.

In summary, existing approaches to implementing neural networks on FPGA soft logic face challenges, particularly when aiming for state-of-the-art accuracy with recent models. TLMAC incorporates concepts from previous BRAM and LUT-based CIM research to realise scalable in-memory computing on current FPGA devices.

\subsection{Neural Network Quantisation} \label{sec: qnn}
Competitive quantisation works endeavour to approximate model data to fewer than 4 bits. Binary neural networks perform extreme quantisation to single-bit values but suffer considerable accuracy loss in both convolution~\cite{liu2018bi, tu2022adabin} and Transformer models~\cite{qin2022bibert}. Conversely, low-bit QNNs maintain high accuracy by two main techniques. First, quantisation itself is performed in high precision, such as scaling~\cite{esser2020learned}, offsetting~\cite{bhalgat2020lsq+} or learned thresholding~\cite{liu2022nonuniform}. Next, mixed precision is common practice, where a minimal portion of the network with high importance is maintained at higher precision, such as skip connections~\cite{choi2018pact} or attention modules~\cite{yao2022zeroquant}.

However, these low-bit, mixed-precision methods are not compatible with most hardware platforms, which compute in integers or bytes~\cite{kim2021bert, jin2022f8net}. While there exists low-bit custom accelerator ASIC chips \cite{sharma2018bit, kim2020exploiting}, such accelerators are less common, less flexible post-fabrication, and not commercially available. Hence, the flexibility of FPGAs motivates its use for accelerating low-bit models.

\section{TLMAC Processing Element}
Multiply-accumulate (MAC) operations make up the bulk of computations in a neural network. Specifically, matrix multiplication constitutes over 90\% of operations in both convolution and Transformer models. Following quantisation, these layers only involve low-bit integer operations. The following sections elaborate on how they are mapped onto lookup tables, forming TLMAC processing elements (PE). Each layer is processed by its dedicated PE, allowing for a dataflow-style accelerator architecture. In this section, we will discuss in general terms the options of using LUTs for MAC computations, before applying the concept to neural networks specifically.

\subsection{Multiply-Accumulate with Lookup Tables} 
\label{sec: mac luts}
An atomic multiply-accumulate operation used frequently in neural networks takes on the following general form in Equation~\ref{eqn: mac operation} when applied to $G$ activations $a_g$ and weights $w_g$, respectively. The result is a partial sum $p$ that, in turn, is accumulated across multiple sequential MAC operations.

\begin{equation} \label{eqn: mac operation}
    p = a_0 \cdot w_0 + a_1 \cdot w_1 + \ldots + a_{G - 1} \cdot w_{G - 1}
\end{equation}

In case of a quantised neural network, weights assume a bit width of $B_w$ and activations $B_a$. The size of the partial sum must be $B_p >= B_w + B_a$, but should be larger depending on $G$ to avoid overflow during accumulation. In practice, the value is determined by the heuristics of partial sums during training.

The invariability of weights $w$ during inference allows us to employ computing-in-memory principles and encode their values into the truth tables stored using LUTs. Hence, a hardware implementation of the MAC operation is left with $G \cdot B_a$ variable input bits and must generate $B_p$ output bits. As we demonstrate our design on an AMD Xilinx Ultrascale+ FPGA, we adopt the LUT-6 lookup table prevalent in their configurable logic blocks~\cite{xilinx-ultrascale-clb}. The LUT-6 maps six input bits to one output bit. In the following paragraphs, we explore two possible methods to arrive at the required MAC mapping using LUT-6.

\subsubsection{Bit-Parallel Implementation}
The number of inputs and outputs of a lookup table can be increased by cascading LUT-6 primitives along the respective dimension. To generate an operation with $B_p$ output bits, a total of $B_p$ LUT-6 can be instantiated and the inputs are shared among them. Since each output bit is generated by a separate LUT, the number of LUTs required scales linearly with the output bit width. For every input bit added to the operation, the size of the truth table and, hence, the number of required LUT-6 doubles. We can express the number of LUT-6 required for an operation with $G$ $B_a$-bit inputs and $B_p$ outputs by Equation~\ref{eqn: luts parallel}.

\begin{equation} \label{eqn: luts parallel}
    N_{\text{lut}} = 2 ^ {G \cdot B_a - 6} \cdot B_p
\end{equation}

The exponential dependency on the number and precision of the input activations renders a pure bit-parallel approach difficult to scale. If we take, as a realistic example, 4-bit wide inputs and 10-bit wide outputs, with $G = 2$ weights involved, we would require an average of $\frac{N_{\text{lut}}}{G} = \frac{40}{2} = 20$ LUTs to store a single weight. With ResNet-18 having over 11.1 million convolution weights, it would require over 200 million LUTs for the MAC operations in convolution layers alone. This figure surpasses the capacity of even the most high-resource FPGA devices available today by orders of magnitude, therefore making this implementation unfeasible.

\subsubsection{Hybrid Bit-Serial Implementation}
In order to tackle the problem of scaling the fan-in, we transform the MAC operation into a input-bit-serial format. We denote a $B_a$-bit activation value $a_g$ in its binary form as $\left\{ a_g^0, a_g^1, \ldots, a_g^{B_a - 1} \right\}, a_g^b \in \{0, 1\}$. The MAC operation in Equation~\ref{eqn: mac operation} can be serialised according to Equation~\ref{eqn: mac operation serial}.

\begin{equation} \label{eqn: mac operation serial}
    p = \sum_{b = 0}^{B_a - 1} 2 ^ b \left( a_0^b \cdot w_0 + a_1^b \cdot w_1 + \ldots + a_{G - 1}^b \cdot w_{G - 1} \right)
\end{equation}

By computing the outer sum iteratively, the required number of input bits of the LUT hardware drops from $G \cdot B_a$ to only $G$. We define the constraint $G \leq 6$, which prevents the operation from exceeding the number of inputs on a LUT-6. LUTs are parallelised only along the output dimension, with the number of LUTs corresponding to the number of output bits. It can be expressed by Equation~\ref{eqn: luts serial}.

\begin{equation} \label{eqn: luts serial}
    N_{\text{lut}} = B_w + \left\lceil \log_2(G) \right\rceil = B_l
\end{equation}

We denote a group of $N_{\text{lut}}$ LUTs cascaded at the output as a \textit{LUT array}. Such an array produces a MAC result of width $B_l$. That is enough to represent the maximum sum of $G$ weights each having a size of $B_w$, in case all input bits are set to~$1$. While the number of LUTs is independent of $B_p$, a $B_p$-bit adder is necessary to accumulate the partial sums $p$ between the serial iterations.

We refer to a set of $G$ weights as a \textit{weight group}. When choosing a $G < 6$, not all inputs of the LUT-6 are utilised. And since LUTs have the ability to implement any arbitrary function, these unused input bits can be repurposed as an additional select signal~$s$. According to Equation~\ref{eqn: number clusters}, LUT arrays can store a total of $N_{\text{clus}}$ different weight groups that can be applied mutually exclusively to the input activation bits $a_g^b$. With the additional selection, $G \cdot N_{\text{clus}}$ weights are stored in $N_{\text{lut}}$ LUTs, significantly decreasing the LUTs-per-weight ratio compared to the bit-parallel implementation. Using our example from above with 4-bit weights and $G = 2$, the LUT-to-weight ratio is $\frac{N_{\text{lut}}}{G \cdot N_{\text{clus}}} = \frac{5}{2 \cdot 16} = 0.16$ -- a more than 100-fold reduction of resources. The hybrid bit-serial implementation makes table lookup-based computing accessible to current FPGA devices.

\begin{equation} \label{eqn: number clusters}
    N_{\text{clus}} = 2 ^ {6 - G}
\end{equation}

Comparing purely bit-parallel to hybrid bit-serial implementation reveals a trade-off between area and latency. Although the first approach achieves single-cycle completion of the MAC operation, it consumes orders of magnitude more resources. Conversely, hybrid operation, takes $B_a$ times more cycles to finish. However, the benefit of bit-serial implementation being scalable far outweighs its cost in terms of reduced speed. More control over this trade-off is provided with parameter $G$. While a higher $G$ increases computational parallelism, more LUTs are demanded to store the weight groups. But matching $G$ with the size of weight groups already present in the neural network is another aspect to be considered, as shown in the next section.

\subsection{Applying TLMAC to Convolution Layers} \label{sec: tlmac application}
\begin{figure}[t]
  \centering
  \includegraphics[width=\linewidth]{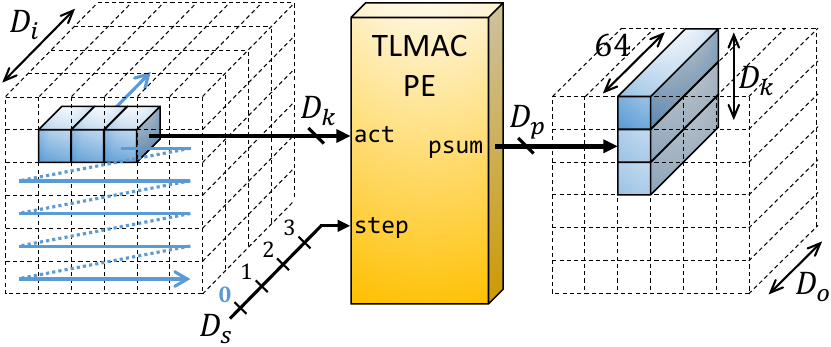}
  \caption{A window of $1 \times D_k$ with $D_k = 3$ values from the input tensor are passed to the TLMAC PE along with the current \texttt{step} index 0 along the $D_s$ dimension. Using all 3 kernel rows in parallel, the PE produces partial sums in three rows of the output feature map, spanning 64 channels. While the first row is fully completed, other ones are pending accumulation with partial sums generated from values in the subsequent rows of the input feature map.}
  \Description{Highlighted values of an input tensor are passed to the TLMAC PE. It produces highlighted values in the output tensor.}
  \label{fig: dimensions}
\end{figure}

The convolution operation in a neural network is used to spatially filter a three-dimensional feature map by sliding a convolution kernel across it. The weights are four-dimensional for the convolution to result in a three-dimensional feature map of same or different size as the input. While the convolution layer involves a deep hierarchy of loops over many dimensions, our focus in this study is primarily on the lowest levels of these loops, as these are the ones deployed on the TLMAC processing element. It is worth noting that dataflow optimisations at higher levels of the hierarchy remain a possibility, orthogonal to the TLMAC implementation, and fall outside the scope of this paper.

Convolution layers use square kernel dimension of size $D_k \times D_k$, with the most common values being 3 or 1. We define a weight group $\mathcal{W} = \left\{ w_0, \ldots, w_{D_k - 1} \right\}$ as a single row of a kernel. With the layer weights having dimensions $D_i$ and $D_o$, representing the number of input and output channels, respectively, the layer contains a total of $D_k \cdot D_i \cdot D_o$ weight groups.

LUT arrays are parameterised with $G = D_k$ to handle the MAC operation of one kernel row, and storing up to $N_{\text{clus}}$ selectable weight groups. To accommodate all weight groups within the layer, the TLMAC PE contains a total of $N_{\text{arr}}$ such units, with the inputs signals shared among them. How exactly the weight groups are allocated to the LUTs will be the topic of Section~\ref{sec: place route}.

Independently, our PE is configured to produce $D_p$ parallel outputs and accumulate the partial sums over $D_s$ sequential inputs, whereby $D_s D_p = D_o D_i D_k$. The exact values for dimensions $D_p$ and $D_s$ are, again, subject to higher-level dataflow adjustments, such as tiling, loop unrolling or loop interchange~\cite{ma2018optimizing}. For this work, we set $D_p = 64 \cdot D_k$. That simultaneously produces partial sums for all kernel rows spanning 64 output channels, based on the same set of $D_k$ input values. Parallelising the input channels $D_i$ is not possible with a single PE since additional LUT inputs would be required. Consequently, the sequential dimension is formed by the input channel dimension and the remaining fraction of $D_o$ in cases where it exceeds 64. Hence, $D_s = D_i \cdot \frac{D_o}{64}$. 

Figure~\ref{fig: dimensions} visualises the input and output shapes and their context within the input and output activation tensors. The $1 \times D_k$ window slides across the input feature map in a row-major order, stepping through the $D_s$ dimension at every window position. At each of those steps, three output feature rows are computed in parallel. Once the last step is finished, the convolution operation for the first row is complete. The other two are buffered in a partial sum memory to be further accumulated when processing the input values in the next row below.

From here on, we denote the format of the weights in terms of parallel and sequential dimensions, as opposed to input and output channels. For our case, the weight tensor is reshaped from $\left[ D_o, D_i, D_k, D_k \right]$ to a tensor of weight groups with size $\left[ D_s, D_p, D_k \right]$.

\section{Hardware Architecture}
\begin{figure}[t]
  \centering
  \includegraphics[width=\linewidth]{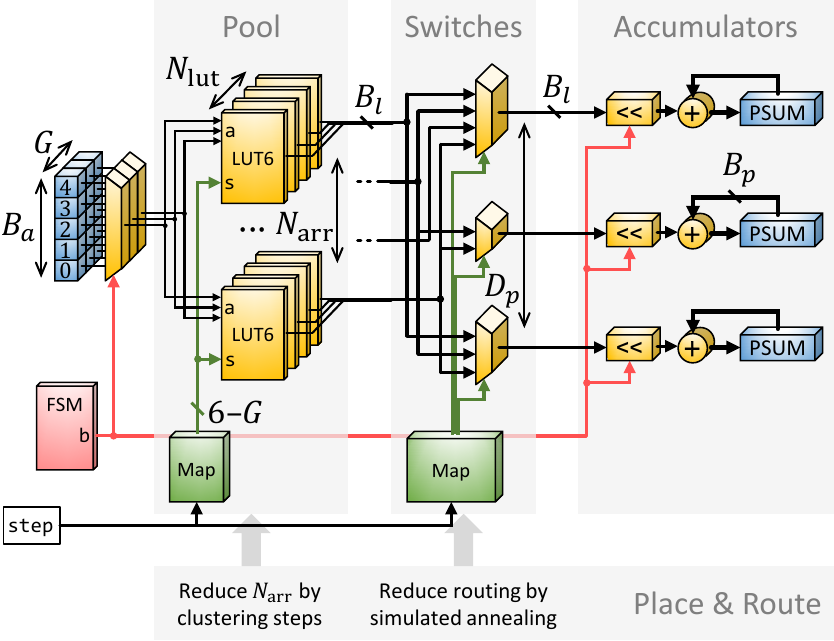}
  \caption{TLMAC processing element consisting of LUT pool, switches and accumulators. $N_{\text{arr}}$ LUT arrays consist of LUTs to generate $N_{\text{lut}}$ output bits. Each LUT array can store $N_{\text{clus}}$ weight groups. Hardwired routing connects LUT outputs to switches that select which MAC result is to be accumulated. The place \& route algorithms target $N_{\text{arr}}$ and routing connections, respectively.}
  \Description{Schematic of TLMAC hardware consisting of logic blocks and arrows indicating signals.}
  \label{fig: tlmac pe}
\end{figure}

In contrast to conventional accelerator designs, the lookup principle eliminates the need for weight transfers into the processing element. Instead, the weights are incorporated in the LUT initialisations during the compilation process and synthesised into the FPGA's bitstream. Therefore, the data interface of the TLMAC PE only consists of input activations $a$, partial sum inputs/outputs $p$, and a \texttt{step} input. The latter indicates the current index along the sequential dimension, with $\texttt{step} \in [0, D_s - 1]$ (see Figure~\ref{fig: dimensions}). A visualisation of the hardware architecture is provided in Figure~\ref{fig: tlmac pe}. The PE is internally controlled by a small state machine that generates control signals, such as the current bit index $b$ of the bit-serial process. The main components include the LUT pool to compute partial sums for all selected weight groups, the switches to select the appropriate LUT result for each output, and the accumulators of partial sums.

For each of the $D_p$ outputs of the processing element, an accumulator stores the partial sum value in a register of length $B_p$. Before the start of the process, the high-precision partial sum values buffered in a block memory outside the PE are loaded into those registers for further accumulation. Activations $a$ are available at the inputs as $B_a$-bit numbers. During each of the bit-serial iterations, one bit $a_g^b$ of each of the $G$ activations is fed into the pool, beginning with the LSB.

The pool's central component is the LUT array as detailed earlier in Section~\ref{sec: mac luts}. The number of those arrays $N_{\text{arr}}$ is determined by the place \& route algorithm. For the LUT count $N_{\text{lut}}$ within each array, refer back to Equation~\ref{eqn: luts serial}. Out of the six inputs of each LUT in the pool, $G$ are used for the activation bits. The remaining ones select the weight group being used. The selection index~$s$ is stored in a read-only mapping memory within the pool that is addressed by the \texttt{step} value, which remains constant throughout the operation. The contents of this memory, i.e. which set of weight groups is selected during each sequential step, is decided during the compilation.

The connections between the output of the LUTs and the accumulators are established through multiplexers. Each of them is linked to a specific subset of LUTs, rather than all $N_{\text{arr}}$ of them. This routing configuration is static and optimised during place \& route. Depending on the \texttt{step} input to the PE, the selection of the LUT array changes. A mapping associating each step with the corresponding selection input for each multiplexer is stored in a read-only memory block.

The MUX output is shifted to the left $b$ times to match the power-of-two scaling of the currently selected input bit. The result is added to the existing high-resolution partial sum stored in the accumulator. After processing all $B_a$ input bits, the result is made available at the outputs of the TLMAC processing element.

\subsection{System-Level Implementation}
Layer control loops that handle the sliding window and consider parameters, such as padding and kernel stride, are implemented on a higher level of abstraction using high-level synthesis (HLS). The TLMAC PEs, written in SystemVerilog, are embedded as blackbox modules and communicate with the rest of the design via streaming data structures (i.e., FIFOs). Those constructs are also used to connect other HLS components, such as layers and ResNet blocks. This dataflow-style architecture, along with dedicated hardware blocks for every layer, allows extensive pipelining and parallelism.

Lookup operations are best suited for low-bit convolution layers. In addition to that, the neural network blocks employ quantisation functions, batch normalisation and skip connections as described in more detail in Section~\ref{sec: quantised neural network}. To faithfully implement the neural network on the FPGA and to replicate the accuracy of the software implementation, those layers involve floating-point computations. We let HLS use the yet untapped DSP resources for that.

\section{Place and Route} \label{sec: place route}
Given a weight tensor of dimensions $[D_s D_p D_k]$ as derived in Section~\ref{sec: tlmac application}, its weight groups of size $D_k$ need be placed onto the LUTs in the TLMAC processing element. This placement is considered an optimisation problem that must adhere to the following hardware constraints:
\begin{enumerate}
    \item Each LUT array can hold a maximum of $N_{\text{clus}}$ weight groups (see Equation~\ref{eqn: number clusters}).
    \item Access to these weight groups is exclusive, meaning that it is not possible to simultaneously use multiple weight groups stored in a single LUT array.
    \item As the selection signal~$s$ is shared, all LUTs use the respective weight group at that index~$s$ for operation.
\end{enumerate}

\begin{figure}[t]
  \centering
  \includegraphics[width=\linewidth]{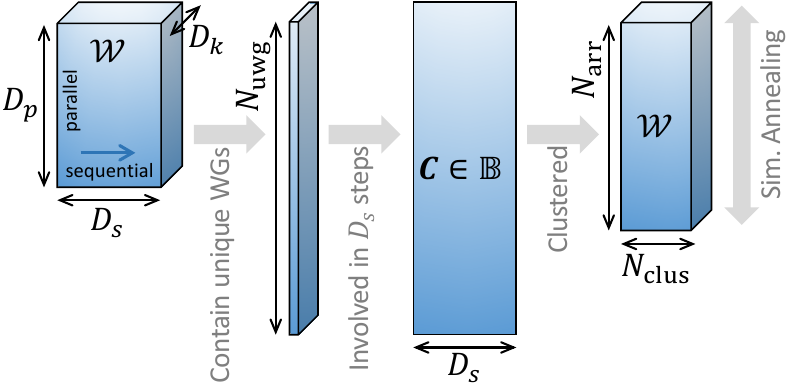}
  \caption{From the weight tensor with parallel and sequential dimensions, the unique weight groups are extracted. A binary assignment matrix $\mathbf{C}$ is derived that shows which of the unique weight groups are involved in every step along $D_s$. After clustering of $D_s$, the horizontal assignment of weights to their respective cluster is fixed. Simulated annealing determines the vertical assignment within $N_{\text{arr}}$ LUT arrays.}
  \Description{Multi-dimensional tensors with dimensions. Arrows show the process step.}
  \label{fig: place route}
\end{figure}

Each phase of our optimisation strategy aims to achieve its own objective: Firstly, $N_{\text{arr}}$ and, hence, the total number of LUTs is to be minimised. This is accomplished by identifying the steps along dimension $D_s$ whose weight groups can be shared and assigned the same index~$s$ within the LUT arrays. Secondly, the wire connections between the LUT pool and the switches are to be reduced by deciding the exact location of the weight groups among the $N_{\text{arr}}$ LUT arrays. That benefits the size of the MUXes, the width of their mapping memory and the level of routing congestion. A favourable optimisation result relies on the quantised weight tensors to have a substantial amount of duplicated weight groups. Our analysis in Section~\ref{sec: weight redundancy} reveals that this holds true for state-of-the-art quantisation techniques. The explanations in the following sections are accompanied by Figure~\ref{fig: place route}.

The result of the Place \& Route process is the assignment of weight groups to the lookup tables in the TLMAC processing element. Based on that, the 64-bit LUT initialisation values are derived and included into the synthesis, which embeds the model weights into the FPGA bitstream.

\subsection{Clustering of Sequential Dimension} \label{sec: clustering}

During each sequential step of the convolution operation, all required MAC results for the respective $D_p$ weight groups must be produced in parallel. That requires each of the weight groups to be placed in the same position of separate LUT arrays, as to satisfy Constraints~(2) and~(3). The weights along dimension $D_s$ of the weight tensor are accessed in sequence. They can be stored within the same LUT array without violating Constraint~(2). However, the number of weight groups per array is limited by Constraint~(1). And since it can be assumed that $D_s > N_{\text{clus}}$, a method is needed to losslessly compress the weights along $D_s$.

Opposed to existing lossy weight clustering techniques that are used for model pruning~\cite{ayinde2019redundant}, we define a \textit{cluster} as a set of \textit{steps}. The union of all weight groups associated with the steps in a cluster are going to be deployed on the LUT arrays under the same index~$s$. With the number of clusters fixed to $N_{\text{clus}}$, our objective is to identify the steps in $D_s$ that share a large number of weight groups, as these groups only need to be stored once within their cluster. That reduces the size of individual clusters and, therefore, the overall LUT utilisation.

We first extract all $N_{\text{uwg}}$ unique weight groups in the weight tensor. A binary assignment matrix $\mathbf{C} \in \mathbb{B}^{D_s \times N_{\text{uwg}}}$ is constructed that highlights, for each step, the weight groups involved in the computation. In the context of clustering, $\mathbf{C}$ can be perceived as $D_s$ samples residing within an $N_{\text{uwg}}$-dimensional binary space. Steps that use a similar set of weight groups are located closely together within that space.

To assign the steps to exactly $N_{\text{clus}}$ clusters, we apply spectral clustering This is a graph-based technique well-suited for high-dimensional data and non-convex cluster shapes. It constructs an affinity matrix between samples by computing a graph of nearest neighbours, followed by a low-dimension embedding of this matrix. The clustering is performed on the components of the eigenvectors in this low-dimensional space. We chose the Cluster QR method as the label assignment strategy to directly extract clusters from eigenvectors without iterations or parameter tuning~\cite{damle2019simple}.

The number of LUT arrays $N_{\text{arr}}$ is determined by the cluster with the maximum number of unique weight groups across all steps within this cluster. The pool of the TLMAC PE contains a BRAM block storing the cluster assignments. It translates the \texttt{step} input to the respective selection value~$s$.

After clustering, the allocation of weight groups to specific selection indices $s$ is established. In which exact LUT array the weight groups are placed is decided during the subsequent routing reduction step.

\subsection{Simulated Annealing for Routing Reduction} \label{sec: simulated annealing}
\begin{algorithm}[t]
    \caption{Simulated Annealing of Routing}
    \label{alg: simulated annealing}
    
    \SetAlgoNlRelativeSize{0}
    \SetAlgoNlRelativeSize{-1}
    \SetNlSty{textbf}{(}{)}
    
    \KwIn{Maximum iterations $I$, Cooling rate $\alpha$}
    \KwOut{Final solution $\mathbf{R_{\text{current}}}$}
    
    Randomly place weight groups and derive $\mathbf{R}$ \;
    $\mathbf{R_{\text{current}}} \gets \mathbf{R}$ \;
    $R_{\text{best}} \gets \text{Count}\left(\mathbf{R_{\text{current}}}\right)$ \tcp*{Equation \ref{eqn: routes count}}
    
    \For{$i \gets 1$ \KwTo $I$}{
        \tcc{Cool down the temperature}
        $T \gets \frac{I}{(i + 1) ^ \alpha}$ \;
        \tcc{Randomly pick cluster and LUT arrays}
        $c \gets \text{RandInt}\left(0, N_{\text{clus}}\right)$ \;
        $e_0 \gets \text{RandInt}\left(0, N_{\text{arr}}\right)$ ;
        $e_1 \gets \text{RandInt}\left(0, N_{\text{arr}}\right)$ \;

        \tcc{Swap to generate neighbouring solution}
        $\mathbf{R_{\text{new}}} \gets \text{Swap}\left(\mathbf{R_{\text{current}}}, c, e_0, e_1\right)$ \;
        $R_{\text{new}} \gets \text{Count}\left(\mathbf{R_{\text{new}}}\right)$ \;

        \tcc{Evaluate new solution}
        \If{$R_{\text{new}} < R_{\text{best}}$ \textbf{or} $\text{Rand}(0, 1) < \exp\left(\frac{R_{\text{best}} - R_{\text{new}} - 1}{T}\right)$}{
            $\mathbf{R_{\text{current}}} \gets \mathbf{R_{\text{new}}}$ \;
            \If{$R_{\text{new}} < R_{\text{best}}$}{
                $R_{\text{best}} \gets R_{\text{new}}$ \;
            }
        }
    }
    \Return{$\mathbf{R_{\text{current}}}$}
\end{algorithm}

The computation of a PE output involves only a subset of the weight groups stored in the LUT arrays. Hence, every one of the $N_{\text{arr}}$ LUT arrays require a physical routing path to all $D_p$ switches. We can leverage this routing sparsity, along with the freedom to move weight groups between LUT arrays, to optimise the overall utilisation of routing resources. For example, if weight groups at distinct indices~$s$ produce MAC results that are used by the same outputs, it is sensible to consolidate them into the same LUT array, as to reuse the wires connecting to the respective switches.

A large number of LUT arrays and parallel outputs, however, results in an intractable number of possible placement combinations for the weight groups, making it an NP-hard problem. Additionally, the solution landscape exhibits a high degree of non-linearity with many local minima. The discrete nature of weight group placement prohibits the application of gradient descent or similar continuous optimisation techniques.

We therefore apply the stochastic method of simulated annealing. Inspired by the annealing process in metallurgy, it is used to find approximate solutions to complex optimisation and search problems. Algorithm~\ref{alg: simulated annealing} starts with randomly assigning weight groups to LUT arrays. The corresponding binary routing matrix $\mathbf{R} \in \mathbb{B}^{N_{\text{arr}} \times N_{\text{clus}} \times D_p}$ indicates if a connection exists between a weight group in the LUTs and any of the $D_p$ outputs. The total number of routes $R$ accounts for connections between the switches and any of the clusters in the LUT arrays:

\begin{equation} \label{eqn: routes count}
    R = \sum_{e=1}^{N_{\text{arr}}}\sum_{p=1}^{D_p} \mathbb{I}\left( \exists c : \mathbf{R}(e, c, p) \neq 0 \right).
\end{equation}

$\mathbb{I}(\cdot)$ denotes the indicator function that yields 1 if the expression is true, and zero otherwise. Over the course of many iterations $I$ (typically $> 10^5$), $\mathbf{R}$ is changed randomly by swapping two weight groups of the same cluster $c$ between arrays $e_0$ and $e_1$. The new solution is accepted if the \textit{energy} of the system, i.e. the total number of routes, has been reduced. Occasionally, worse solutions are accepted to escape local minima. The likelihood of this happening depends on the \textit{temperature} $T$ that is scheduled to be reduced exponentially over time. We chose a temperature decay $\alpha = 1.4$.

Our tests show that this process can achieve a routing reduction of up to 50\%, starting from a random weight group assignment. From the final assignment of weight groups to the LUT arrays, the static wire connections can be extracted. They are coded into the design files during compilation and included into the bitstream during hardware implementation.

\section{Experimental Results}
TLMAC capitalises on quantisation strategies developed separately and maps pre-trained model parameters onto the LUTs, guaranteeing equivalence between FPGA and software computations. Consequently, the accuracy reported below is consistent with the original quantisation papers.

The primary benchmark of those quantisation works is ImageNet, a complex image classification dataset consisting of 1000 classes that emerged as the de facto standard for evaluating practical computer vision algorithms. Previous works on soft-logic computing frequently lack the scalability to support the model sizes necessary for ImageNet classification~\cite{andronic2023polylut, umuroglu2020logicnets}. Among the existing solutions, only LUTNet~\cite{wang2019lutnet} and Logic Shrinkage~\cite{wang2022logic} have reported results on this dataset, albeit with certain constraints.

Our first set of experiments, hence, align with the principles established by LUTNet, which only unrolled and deployed a single ResNet block onto the physical FPGA. For fair comparison we adopt their approach and report the area and accuracy metrics using the same ResNet block quantised with N2UQ at varying bit widths. In subsequent experiments, we demonstrate the capability of TLMAC to accommodate all ResNet basic blocks on the FPGA, thereby surpassing the limitations of prior methods.

We conducted our implementation on the AMD Xilinx Virtex UltraScale+ XCVU13P FPGA, operating at a clock frequency of 200~MHz. We employed the default strategies provided by the Vivado toolchain for both synthesis and implementation. Energy consumption was estimated using the default settings in Vivado.

\subsection{Quantised Neural Network Models} \label{sec: quantised neural network}
We implement Nonuniform-to-Uniform Quantisation~\cite{liu2022nonuniform} as it is currently the most competitive low-bit model, surpassing full-precision accuracy using only 3-bits. N2UQ combines the improved representational capabilities of non-uniform quantisation with the low computation cost of uniform quantisation using a learnable thresholding function. This applies to convolution layers, while batch normalisation, quantisation and activation functions remain in floating-point precision. Furthermore, the first and last layers are maintained at full precision as in common practice.

To optimally leverage the low-precision capabilities of FPGAs, our implementation focuses on the basic blocks within ResNet-18, whose convolution layers are deployed on TLMAC processing elements. For floating-point layers, we take advantage of the abundant, still unutilised DSP slices to compute these complex operations. Additionally, the first convolution and the fully connected layer are offloaded to the host PC, capitalising on high-speed floating-point processing capabilities. These design partitioning decisions ensure efficient utilisation of available resources.

\subsubsection{Weight Redundancy in Quantised Neural Networks} \label{sec: weight redundancy}
\begin{figure}[t]
  \centering
  \includegraphics[width=\linewidth]{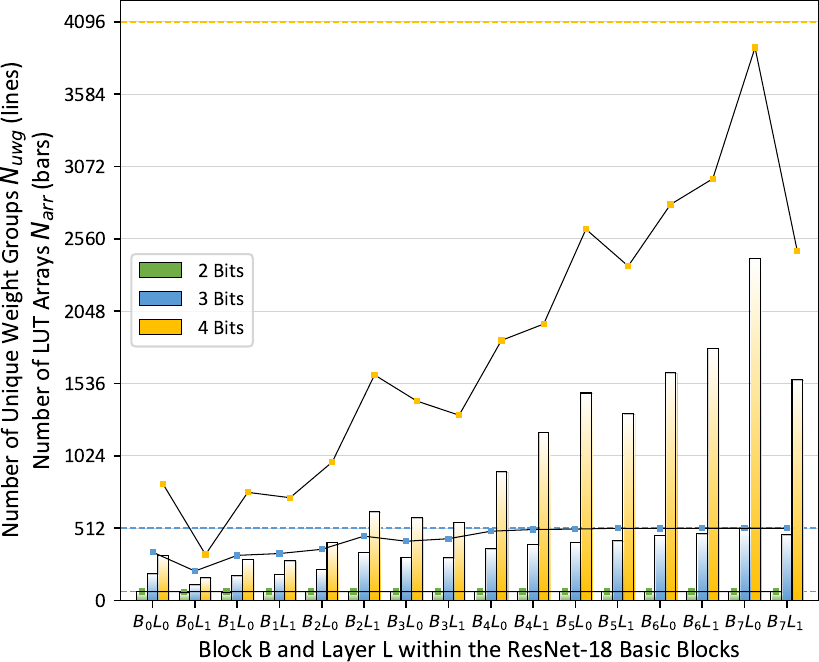}
  \caption{Lines show the number of unique weight groups in convolution layers within ResNet-18's basic blocks. The theoretical maximum of weight groups based on the weight bit width and kernel size is given by the dashed horizontal lines, respectively. The result of the clustering are $N_{\text{arr}}$ LUT arrays per layer represented by the bars.}
  \Description{Bar graph showing number of unique weight groups for every layer.}
  \label{fig: plot logic density}
\end{figure}

Neural networks that were quantised down to a low bit width exhibit a considerable degree of weight redundancy. Even when examining groups of weights $\mathcal{W}$, it becomes evident that layers are made up of a relatively small number of unique weight groups. The lines in Figure~\ref{fig: plot logic density} illustrate the total number unique weight groups by layer in the quantised ResNet-18 basic blocks. The theoretical maximum number of unique weight groups, depending on the bit width and assuming kernel size $D_k = 3$, is represented by dashed horizontal lines. Even for layers that reach this maximum, unique weight groups make up only a small portion of the total number of parameters in the layer, often fewer than 5\% and less for bigger layers.

The small number of unique weight groups has the following implications: Firstly, embedding all weight groups in LUTs as part of their truth table becomes attainable. Secondly this sparsity results in each unique weight group participating in numerous MAC operations within the layer. Therefore, weight groups need to be accessible at the time they are required for convolution.

With TLMAC, we are able to leverage the weight redundancy to achieve scalability, with place \& route algorithms ensuring the correct access to weight groups. The next sections will analyse in detail how the number of unique weight groups translates into LUT utilisation and power consumption.

\subsection{Area and Energy Efficiency}
\subsubsection{Logic Density}
We apply clustering to the unique weight groups as described in Section~\ref{sec: clustering} in order to maximise the \textit{logic density}, while working around the constraints of the hardware architecture. Logic density is a metric used in~\cite{wang2019lutnet} that is defined as "the number of LUTs required to construct a network able to achieve a particular test accuracy". In the context of clustering in TLMAC, it also describes the number of unique weight groups stored per LUT array. Recall that Equation~\ref{eqn: number clusters} gives the maximum number of weight groups per LUT array as $N_{\text{clus}}$, which is 8 for $3\times3$ convolution layers.

The coloured bars in Figure~\ref{fig: plot logic density} illustrate the count of LUT arrays $N_{\text{arr}}$ for each $3\times3$ convolution layer within the basic blocks. The actual logic density for each layer can be derived by taking the ratio $\frac{N_{\text{uwg}}}{N_{\text{arr}}}$. Additionally, the overall logic density is obtained by dividing unique weight groups in all blocks by the total number of LUT arrays. For bit widths 2, 3 and 4, the overall logic densities are 1.01, 1.30 and 1.86, respectively.

Although these values fall short of the theoretical maximum of 8, clustering significantly reduces the demand for LUT resources, with reductions of up to 23\% for 3-bit models and 46\% for the 4-bit models. Particularly higher-precision networks benefit due to their high logic resource requirements. Additionally, not utilising the full capacity of the LUTs provides needed flexibility for routing optimisation to be effective. A further reduction of logic density is prevented by the hardware constraints listed earlier. This is especially evident in the 2-bit model, where nearly every weight group is utilised for MAC operations across all steps. As a result, their MAC results must remain continuously accessible at the pool outputs, making them unsuitable for mutually exclusive access.

\subsubsection{Routing Reduction}
\begin{figure}[t]
  \centering
  \includegraphics[width=\linewidth]{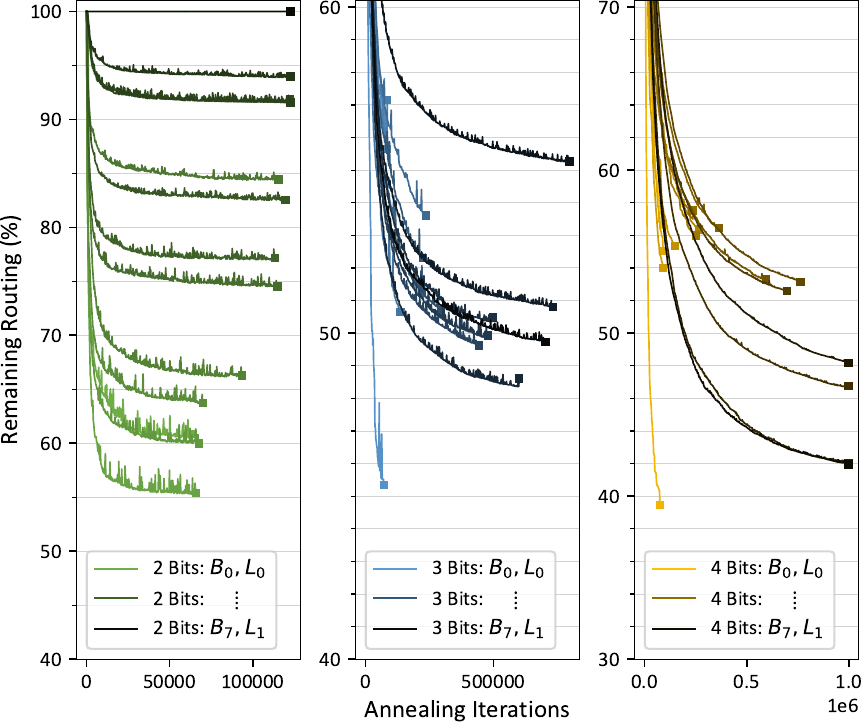}
  \caption{Relative routing reduction obtained through simulated annealing for 2, 3 and 4-bit models. Darker lines represent larger layers in the blocks towards the end of the ResNet-18 model.}
  \Description{Three plots showing exponentially decreasing lines.}
  \label{fig: plot routing reduction}
\end{figure}

The allocation of weight groups to different LUT arrays within in same cluster impacts the number of routing connections between the pool and the switches as detailed in Section~\ref{sec: simulated annealing}. Figure~\ref{fig: plot routing reduction} visualises the simulated annealing progress for 2, 3 and 4 bit models, starting from a random assignment of weight groups to LUT arrays. For each layer in the ResNet-18 basic blocks, we plot the percentage of remaining connections against the annealer iterations. Thereby, darker lines belong to the larger layers towards the end of the models. Is it clearly visible how simulated annealing is capable of escaping local minima in pursuit of reaching a global optimum.

To optimise runtime, the total number of iterations is determined for each layer in proportion to its initial number of connections after the random assignment. Hence, the square markers in Figure~\ref{fig: plot routing reduction} not only show the final reduction along the vertical axis, but also indirectly reflect the absolute number of connections along the horizontal axis of the plot.

Early smaller layers in the models generally have fewer connections, going hand in hand with their lower number of LUT arrays. A high degree of reduction down to less than 50\% is achieved for very early and late layers. An exception are the last layers of the 2-bit model, that requires almost complete connectivity between pool and switches.

These experiments reveal the significant influence of weight group placement among LUT arrays on the number of wires needed to connect the pool outputs and the switches. Simulated annealing proved to be an effective tool for addressing this complex optimisation problem.

\subsubsection{Comparison with Prior Works}
\begin{table*}[t]
\begin{threeparttable}
  \caption{Classification accuracy, area requirements and power consumption for LUTNet, LogicShrinkage and different bit width of TLMAC. ResNet-18 models were trained to classify the ImageNet dataset. Results were obtained by implementing the sixth, 256-channel block of these models on hardware. Deltas $\Delta$ are with regards to LogicShrinkage, the current state of the art.}
  \label{tab: comparison}
  \renewcommand{\arraystretch}{1.3}
  \begin{tabular*}{\textwidth}{@{\extracolsep{\fill}}*{11}{c}}
    \toprule

    \multirow{2}{*}{Architecture} & \multirow{2}{*}{Bits} & \multicolumn{2}{c}{Accuracy} & \multicolumn{2}{c}{Block Area (post-syn.)} & \multicolumn{3}{c}{Block Area (post-impl.)}      & \multicolumn{2}{c}{Block Power (W)} \\
                                                            \cline{3-4}                    \cline{5-6}                                  \cline{7-9}                                        \cline{10-11}
                                  &                       & \% & $\Delta$ (pp)           & LUTs & $\Delta (\times \downarrow)$        & LUTs & $\Delta (\times \downarrow)$ & BRAM       & Dynamic & Static                    \\

    \midrule
    
    LUTNet \cite{wang2019lutnet}        & 1 & 54.87          &  1.47 & \multicolumn{1}{r}{1 840 666} &  0.4 & \multicolumn{2}{c}{---\tnote{b}}     &    --- & \multicolumn{2}{c}{---} \\
    LogicShrinkage \cite{wang2022logic} & 1 & 53.40          &  0.00 & \multicolumn{1}{r}{  690 357} &  1.0 & \multicolumn{1}{r}{665 720}   &  1.0 &    --- & \multicolumn{2}{c}{---} \\
    
    \hline
    
    \multirow{3}{*}{TLMAC}              & 2 & 69.42\tnote{a} & 16.02 & \multicolumn{1}{r}{   54 973} & 12.6 & \multicolumn{1}{r}{ 54 716}   & 12.2 &  79.5  & 0.6 & 3.0                        \\
                                        & 3 & 71.94\tnote{a} & 18.54 & \multicolumn{1}{r}{  112 000} &  6.2 & \multicolumn{1}{r}{110 391}   &  6.0 &  97.0  & 1.0 & 3.0                        \\
                                        & 4 & 72.88\tnote{a} & 19.48 & \multicolumn{1}{r}{  187 908} &  3.7 & \multicolumn{1}{r}{186 435}   &  3.6 & 103.5  & 3.1 & 3.0                        \\

    \bottomrule
  \end{tabular*}
  
  \begin{tablenotes}
    \item[a] Accuracy as reported in \cite{liu2022nonuniform}
    \item[b] Design could not fit onto target device
 \end{tablenotes}
\end{threeparttable}
\end{table*}

\begin{figure}[t]
  \centering
  \includegraphics[width=\linewidth]{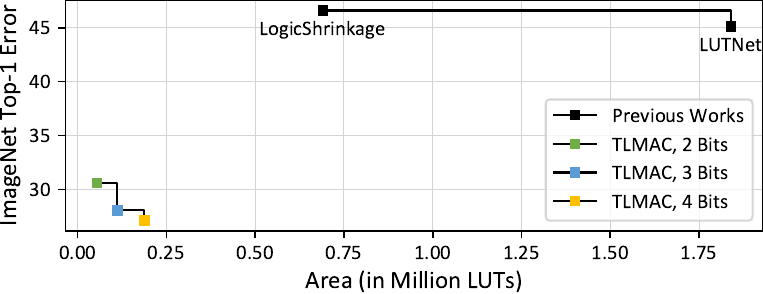}
  \caption{Area-accuracy trade-off for TLMAC and previous works LUTNet~\cite{wang2019lutnet} and LogicShrinkage~\cite{wang2022logic}. TLMAC shifts the Pareto frontier further inward as lower error can be achieved with significantly smaller area.}
  \Description{Point plot showing the location of TLMAC and previous works in the area-error space.}
  \label{fig: plot comparison}
\end{figure}

To put the capabilities of the TLMAC hardware architecture and compilation algorithms into perspective, we conduct comparisons with state-of-the-art soft-logic computing studies that reported findings on ImageNet classification~\cite{wang2022logic, wang2019lutnet}. Following their experimental approach, we synthesised and implemented the sixth ResNet-18 block. This block comprises two convolution layers, two batch normalisation layers and activation functions, each having 256 channels. Additionally, the skip connection adds the block inputs with the output of the second batchnorm layer.
TLMAC processing elements are instantiated for both convolution layers. As shown in Table~\ref{tab: comparison}, the area utilisation, consisting of LUTs and BRAMs, scales nearly linearly with the precision of weights and activations. At higher bit widths, the increased area utilisation is not solely attributed to the higher quantity of LUT arrays $N_{\text{arr}}$. As explained in Section~\ref{sec: mac luts}, the size of each LUT array $N_{\text{lut}}$ also expands to produce high-precision MAC results. The remaining layers are kept in floating-point precision. 12 DSP slices are used for those operations, independent of the quantisation bit width.

Comparing accuracy and area metrics with previous research reveals the strengths of the TLMAC approach. Because it does not require custom training, we can make use of state-of-the-art quantisation works to achieve supreme accuracy with neural networks that, in some cases, even exceeds their floating-point baseline~\cite{liu2022nonuniform}. The accuracy naturally depends on the precision or weights and activations. LUTNet and LogicShrinkage focus on the implementation of binary neural networks, which cannot achieve the same levels of accuracy as their multi-bit counterparts. One might expect that fewer logic resources would be needed in the case of binary neural networks. However, Table~\ref{tab: comparison} shows that TLMAC required over $12\times$ fewer LUT resources for 2-bit, and $3\times$ fewer for 4-bit networks.

The linear relationship is also evident between power and bit width, consistent with the findings of~\cite{wang2019lutnet}, who previously established a strong correlation between area utilisation and power. Although previous works did not report their power consumption for ImageNet models, we can infer from this linear relationship that TLMAC's power consumption is likely to be more efficient. This positions TLMAC outperforms previous works on all metrics.

Another representation of the implementation results is offered in Figure~\ref{fig: plot comparison}. By plotting the classification error rate against the area efficiency, we can illustrate trade-offs. It is generally expected that an increase in area efficiency leads to a decrease in accuracy when moving along the Pareto frontier. Both previous works and TLMAC align with this expectation. However, TLMAC significantly shifts the Pareto frontier inward, reflecting the substantial improvements it brings over prior works.

\subsection{Complete Network Implementations}
Leveraging cutting-edge quantisation schemes, it is possible to achieve floating-point classification accuracy with just 3 bits~\cite{liu2022nonuniform}. TLMAC shows how this can lead to implementing the most compute-intensive layers of these models entirely using soft logic in FPGAs. This allows TLMAC to outperform previous methods that were either constrained by device resource, and/or a low classification accuracy due to the use of binary neural networks. To push this further, we attempted to implement the even larger 4-bit model.

\subsubsection{Implementation Results}
\begin{figure}[t]
  \centering
  \includegraphics[width=\linewidth]{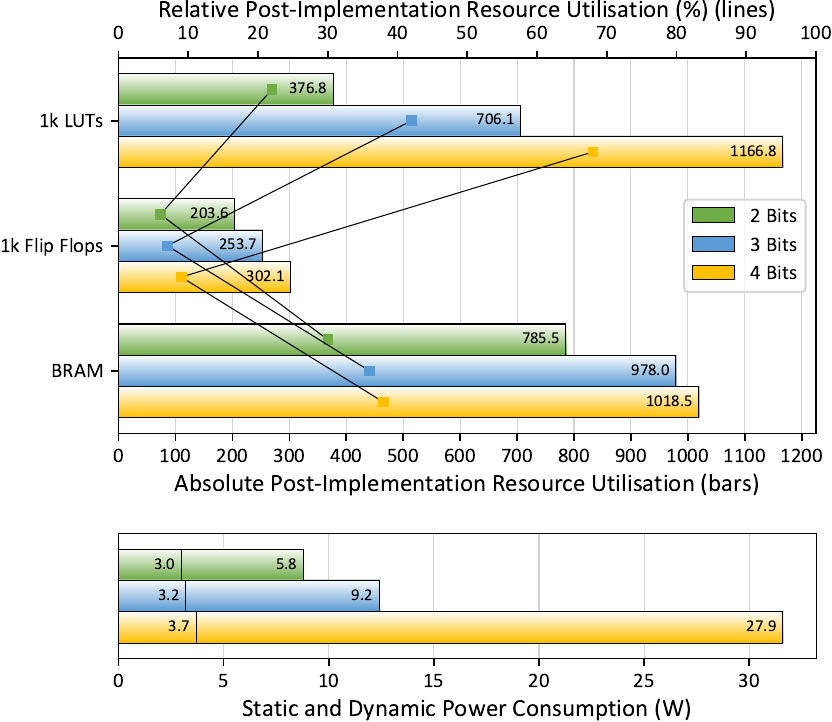}
  \caption{Above: Absolute (bars) and relative (lines) utilisation of different resources for 2, 3 and 4-bit implementations of all ResNet-18 basic blocks. Below: Estimations of static and dynamic power consumption.}
  \Description{Line and bar graphs for resource utilisation and power.}
  \label{fig: plot complete network}
\end{figure}

Figure~\ref{fig: plot complete network} visualises the metrics post-implementation for all ResNet-18 basic blocks. The figure reveals the linear relationship between the model's precision and the utilisation of LUTs, flip flops, and BRAMs. This observation is in line with the data gathered from single-block implementations.

The lower plot in Figure~\ref{fig: plot complete network} provides estimates of the static and dynamic power consumption, obtained through the Vivado power analysis tool. The static power consumption remains stable at around 3~W for 2 and 3-bit implementations. In line with our prior observations, the dynamic power consumption exhibits a nearly linear increase in tandem with the utilisation of logic resources.

The device maps in Figure~\ref{fig: device maps} visualise the physical locations of hardware modules in the FPGA. Thereby, each distinct colour corresponds to one of the eight basic blocks in ResNet. For bits 2 and 3, the placement was entirely delegated to the Vivado tool, without imposing additional constraints. While placement eventually succeeded and timing was met, it can be seen from maps (a) and (b) that layers within the same blocks are frequently spread out across different logic regions (SLRs). Despite connectivity between layers only consisting of FIFO buffers and handshake signals, a more optimal placement would likely locate the layers in close proximity to one another, aligning with the data flow.

To summarise, TLMAC has facilitated the complete implementation of the 3-bit ResNet-18 in soft logic, capable of achieving full-precision accuracy on ImageNet.

\subsubsection{Pushing the Limits}
\begin{figure}[t]
  \centering
  \includegraphics[width=\linewidth]{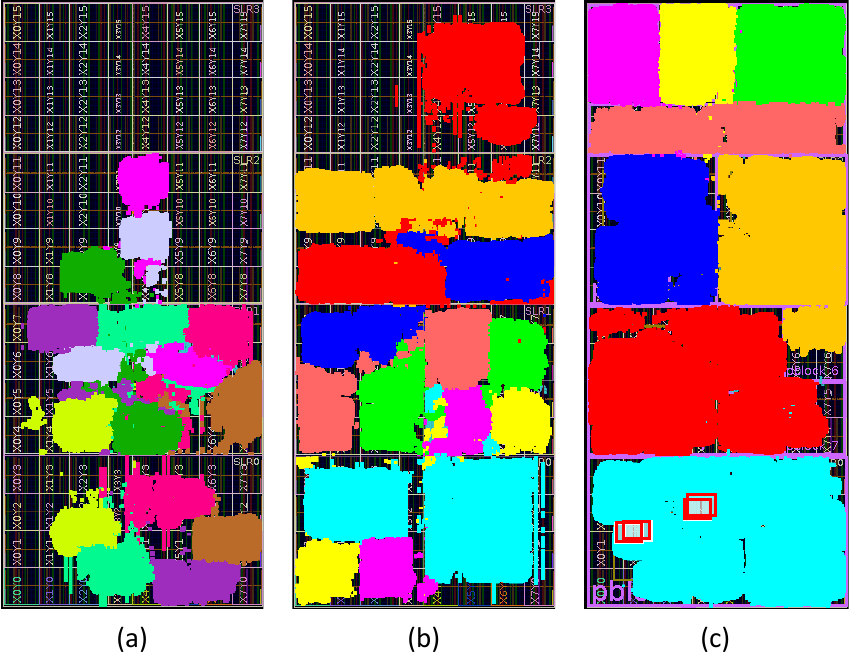}
  \caption{Device maps that highlight the eight basic blocks of ResNet in different colours. For 2 bits (a) and 3 bits (b), the placement was determined by Vivado alone. Placement for 4 bits (c) required constraining the location of the blocks. Still, routing was unsuccessful in the areas highlighted in red.}
  \Description{Device map of the FPGA with each basic block highlighted in a different colour.}
  \label{fig: device maps}
\end{figure}

To explore the limits of current FPGAs, we attempted the additional implementation of a 4-bit ResNet model on the Xilinx XCVU13P, one of the largest available FPGA devices to date. From our experiences with lower-precision implementations, we knew that explicitly laying out a floorplan for the hardware modules on the physical device could enhance our chances of success. Figure~\ref{fig: device maps}(c) shows purple bounding boxes (PBlocks) for each ResNet block that confine hardware resources to the specific areas. Our approach aimed to prevent the allocation of blocks across multiple SLRs and to ensure that the per-PBlock LUT utilisation is below 80\% for any given ResNet block.

These design recommendations resulted in the successful synthesis and placement of the 4-bit network. Furthermore, routing for nets in ResNet blocks 1 to 7 was completed without issues. However, block 8 presented a challenge, with four regions experiencing routing congestion of level 5 that Vivado could not resolve. Those regions are highlighted as red boxes in Figure~\ref{fig: device maps}(c).

In conclusion, TLMAC managed to reduce the logic requirements of the 4-bit model sufficiently to fit onto the device. The routing demands, however, are exceeding the limits of what today's FPGAs can handle.

\section{Conclusion}
In this paper, we presented \textit{table-lookup multiply-accumulate} (TLMAC), a framework that enables the scalable deployment of quantised neural networks on the soft logic on FPGAs. By exploiting recent advances in network quantisation, TLMAC achieves high accuracy using LUT-based computing that alleviates the overhead costs of repeated weight transfers to compute units in more conventional paradigms. TLMAC employs a hybrid bit-serial/bit-parallel approach to maintain the scalability of multi-bit MAC operations. Furthermore, optimisation of weight group placement and routing through clustering and simulated annealing, respectively, enables weights to be densely packed into logic. As a result, it achieves full-precision accuracy on ImageNet of 71.8\%, while fitting the entire model in the soft logic of a single, off-the-shelf FPGA device.

\begin{acks}
This work was supported by the Singapore Government’s Research, Innovation and Enterprise 2020 Plan (Advanced Manufacturing and Engineering domain) under Grant A1687b0033 and further by the National Research Foundation Singapore, under its Quantum Engineering Programme 2.0 (National Quantum Computing Hub, NRF2021-QEP2-02-P01) and funding from Agency for Science, Technology and Research (\#21709) and IAF (IAF311014Q).
\end{acks}

\balance
\bibliographystyle{ACM-Reference-Format}
\bibliography{fpga_2024}

\end{document}